\documentclass[11pt]{article}

\usepackage{scicite}

\usepackage{float}
\usepackage{outlines}
\usepackage{amsmath}
\usepackage{amssymb}
\usepackage{authblk}
\usepackage{graphicx}
\usepackage[braket, qm]{qcircuit}
\usepackage{url}
\usepackage{scalerel}
\DeclareMathOperator*{\barr}{\scalerel*{+}{\textstyle\sum}}

\renewcommand{\i}{i}

\usepackage[style=numeric,bibstyle=numeric]{biblatex}
\renewbibmacro{in:}{}
\newcounter{lastnote}

\title{Knapsack Problem variants of QAOA \\ for battery revenue optimisation}

\author{P. Dupuy de la Grand'rive$^{1}$, J.-F. Hullo$^{1}$}
\affil{$^{1}$EDF Energy R\&D UK centre, Hove, United Kingdom}

\date{}

\begin{document}

\maketitle

\baselineskip=20pt

\begin{abstract}
 
\normalsize
 \baselineskip=20pt
We implement two Quantum Approximate Optimisation Algorithm (QAOA) variants for a battery revenue optimisation problem, equivalent to the weakly NP-hard Knapsack Problem. Both approaches investigate how to tackle constrained problems with QAOA. A first 'constrained' approach introduces a quadratic penalty to enforce the constraint to be respected strictly and reformulates the problem into an Ising Problem. However, simulations on IBM's simulator highlight non-convergent results for intermediate depth ($ p\leq 50$). A second 'relaxed' approach applies the QAOA with a non-Ising target function to compute a linear penalty, running in time $O(p(\log_2 n)^3)$ and needing $O(n \log n)$ qubits. Simulations reveal an exponential improvement over the number of depth levels and obtain approximations about $0.95$ of the optimum with shallow depth ($p \leq 10$).


\end{abstract}
\baselineskip=20pt

\pagebreak
\tableofcontents

\pagebreak

\section*{Introduction}
Among recent quantum algorithms, the Quantum Approximate Optimisation Algorithm (QAOA) has created great expectations for combinatorial optimisation problems \cite{Farhi2014AAlgorithm}. This hybrid quantum-classical variational algorithm is today an active field of research. Particularly active topics of investigation are the optimisation of the angles $\beta, \gamma$ in the computation, the study of driver Hamiltonian, as well as the experimentation of the performance of the algorithm for shallow depths. 

However, the set of problems that have been explored with a QAOA approach remains limited. Recent research have investigated unconstrained problems like MAX-CUT \cite{Zhou2018QuantumDevices} \cite{Wang2018AnAlgorithm} or Max Independent Set (MIS) \cite{MIS}. But to the knowledge of the authors, no paper have yet been published to apply QAOA to non-Ising objective function. As an example, the QAOA algorithm implemented in the Aqua library on Qiskit \cite{Qiskit} currently only supports Ising problems. 

Since adiabatic theorem applies for any time-dependent Hamiltonian \cite{adiabatictheorem}, QAOA is supposed to be working with any objective function. In this paper, we investigate the performance of the QAOA algorithm for non-Ising objective function: : the revenue optimisation of a battery offering services to the power grid. As highlighted in section \ref{sec:relaxed}, this approach enables to integrate constraints with a penalty uncomputable with an Ising formulation. The results of the 'relaxed' approach illustrate the accuracy of this approach and contribute to broaden the range of problems studied with a QAOA approach. 

\newpage
\section{Scheduling optimisation of electricity storage systems}

Electricity storage systems are assets that offer flexibility to process the large-scale integration of renewable energy sources in the power grid. When investing in a large fleet of energy storage systems (batteries for instance), companies seek to evaluate the most likely optimal return on investment of different battery types and specifications, knowing that revenues from batteries come from different type of services sold to the grid. Optimisation of revenues over the battery life cycle hence requires taking into account the returns on these markets, based on price forecast, as well as expected battery degradation over time.

Stimulated by the increasing number of electricity storage assets, different approaches to solve this combinatorial problem using optimisation on classical computers have been published, usually using Mixed Integer Linear Programming \cite{jin2013optimizing}, \cite{liu2015bidding}, \cite{terlouw2019multi}.  We are investigating in this paper how recent advances in quantum optimisation could be adapted to tackle this type problem, and we start in this section by presenting our mathematical model.

\subsection{Energy storage systems (batteries) and electricity markets}

From a modelling perspective, batteries are energy storage systems which transacts energy with the grid by first absorbing electricity (charge) and then supplying electricity (discharge). As any battery, they are subject to ageing, also known as degradation. That degradation affects the battery health and efficiency, and is usually characterised in its simplest form by manufacturers as a maximum charge/discharge cycle numbers to be reached.

The two main electricity markets for batteries are the capacity (volume) and frequency response (frequency) markets, which helps the grid operator to deliver enough (capacity) and good quality (frequency) electricity. On these markets, offer and demand are organised by a regulator that asks each supplier to choose a market in advance, for each time window. Then, the batteries operator will charge and discharge in the network depending on pre-agreed contracts. The supplier makes therefore forecasts on the return and the number of charge/discharge cycles for each time window to optimise its overall return. Since the performance of a battery decreases while it is used, choosing the best cash return for every time window one after the other, without considering the degradation, does not lead to an optimal return over the lifetime of the battery, i.e. before the number of cycles is reached.

\subsection{Problem modelling}

We model the problem as follows: considering two markets $M_1$ and $M_2$, during every time window (typically a day), the battery operates on either one or the other market, for a maximum of $n$ time windows (typically 10 years). Every day is considered independent and the intraday optimisation a standalone problem: every morning the battery starts with the same level of power so that we don't consider charging problems. Forecasts on both markets being available for the $n$ time windows, we assume known for each time window $t$ (day) and for each market:

\begin{itemize}
  \item the daily returns $\lambda_1^t$ and $\lambda_2^t$
  \item the daily degradation, or \textit{health cost} (number of cycles), for the battery $c_1^t$ and $c_2^t$
\end{itemize}

We want to find the optimal schedule, i.e. optimise the life time return with a cost less than $C_{max}$ cycles. We introduce $d = \max_t \{c_1^t, c_2^t\}$. Data show that $d = O(1)$ and it will happen that we ignore it in the calcul of the complexity.

We introduce the decision variable $z_t, \forall t \in [\![1,n]\!]$  such that $z_t = 0$ if the supplier chooses $M_1$ and $z_t = 1$ if he chooses $M_2$, with every possible vector $z=[z_1, ...,z_n]$ being a possible schedule. The previously formulated problem can then be expressed as:

\begin{equation}
  \max_{z \in \{0,1\}^n} \quad \sum_{t = 1}^{n}
  (1-z_{t})\lambda_{1}^{t} + z_{t}\lambda_{2}^{t}
\end{equation}

\begin{equation}
  s.t. \hspace{5mm} \sum_{t=1}^n \big[(1-z_{t})c_1^{t} + z_{t}c_2^{t}\big]  \leq C_{max}
\end{equation}

\subsection{Knapsack Problem formulation}

For a time window $t$, there are 4 situations that need to be dealt with:
\begin{enumerate}
  \item $\lambda_1^t \geq \lambda_2^t$ and $c_1^t \leq c_2^t$. In this case it is trivial that we choose $M_1$.
  \item In the reverse case ($\lambda_2^t \geq \lambda_1^t$, $c_2^t \leq c_1^t$), we will obviously choose $M_2$.
  \item $\lambda_2^t \geq \lambda_1^t$ and $c_2^t \geq c_1^t$. In this case, the choice of $M_2$ is equivalent to choosing an object of value $(\lambda_2^t - \lambda_1^t)$ and of weight $(c_2^t - c_1^t)$.
  \item The case ($\lambda_1^t \geq \lambda_2^t$, $c_1^t \geq c_2^t$) is the contrary of the previous one.
\end{enumerate}

Situations 1 and 2 need no optimisation because we already know what to choose. While situation 3 and 4 are symmetrical, we can consider we are always in case 3. We introduce $p_t = (\lambda_2^t - \lambda_1^t)$ and $w_t = (c_2^t - c_1^t)$. The problem is thus:

\begin{equation}
  \max_{z_t \in \{0,1\}^n} \hspace{2mm} \sum_{t=1}^n p_tz_t \end{equation}

\begin{equation}
  s.t. \hspace{4mm} \sum_{t=1}^n w_tz_t \leq C'_{max} = C_{max} - \sum_{t=1}^n c_1^t
\end{equation}

This formulation illustrates that this problem is equivalent to a Knapsack Problem, a standard form of problem, known to be weakly NP-hard. A classical computing approach deals with dynamic programming, and achieves a $O(nC'_{max})$ complexity. We refer to \cite{karp1972reducibility} for further information on the Knapsack problem.

An adiabatic quantum computing approach has recently been proposed to the Knapsack Problem \cite{Coffey2017AdiabaticProblem}. The author introduces a complex Hamiltonian that enables to take into account the constraint. The paper was published originally for a quantum annealer. On quantum annealer the required evolution time is dictated by the inverse square of the spectral gap, the minimum between the ground and first excited state and it outputs the optimum. We use here the same Hamiltonian for a QAOA approach. We can therefore explore the performances of the algorithm depending on $p$, delivering an approximation with a short depth, compared to the spectral gap, while the adiabatic approach outputs the optimum in a longer time.

\newpage
\section{Quantum Approximate Optimisation Algorithm (QAOA)}

This section aims to give a quick overview of the QAOA algorithm and present briefly the choices we made for the implementation. 

\subsection{Overview}

Many problems can be framed as combinatorial optimisation problems. These are problems defined on n-bit binary strings $z = z_1z_2...z_n, z_i \in \{0,1\}$ where the goal is to determine the a string that maximises yhe objective function $f$.

With $\ket{z} = \ket{z_1...z_n}$ the quantum encoding of $z$, we treat first the objective function as an operator $C$ such that $C\ket{z} = f(z)\ket{z}$ and we define the associated unitary operator $U(C,\gamma) = e^{-i \gamma C}$. Given $\sigma_t^x$ the operator $\begin{pmatrix}0 & 1\\ 1 & 0\end{pmatrix}$ applied to the t-th qubit, we define as well the operator $B$, the sum of all single qubit operators $\sigma^x$, $B =\sum_{t=1}^n \sigma_t^x$ and the associated operator $U(B,\beta)=e^{-i \beta B}$. We set initially the qubits in a uniform superposition of all states:
\begin{equation}
  \ket{s} = \frac{1}{\sqrt{2^n}}\sum_{z\in\{0,1\}^n} \ket{z}
\end{equation}

Given $2p$ angles $\beta \equiv \beta_1,...,\beta_p$ and $\gamma \equiv \gamma_1,...,\gamma_p$, we define the quantum state:
\begin{equation}
  \ket{\beta,\gamma} \equiv U(B,\beta_p) U(C,\gamma_p) .....U(B,\beta_1)U(C,\gamma_1)\ket{s}
\end{equation}

This state can be computed on a quantum computer with a circuit of size growing linearly with $p$. $F_p(\beta,\gamma) \equiv \bra{\beta,\gamma}C\ket{\beta,\gamma}$ is the expected value of $C$ for this state. Let $M_p = \max_{\beta,\gamma} F_p(\beta,\gamma)$. The quantum adiabatic theorem states:
\begin{equation}
  \lim_{p\to\infty}M_p = \max_{z \in \{0,1\}^n}  C(z)
\end{equation}

The idea of the QAOA \cite{Farhi2014AAlgorithm}, an algorithm derived from the Quantum Adiabatic Algorithm \cite{farhi2001quantum}, is to compute the state $\ket{\beta,\gamma}$. With an accurate choice of angles and depth $p$, the value of the output state $C\ket{\beta,\gamma}$ will approximate $\max_{z \in \{0,1\}^n}  C(z)$. We refer the reader to \cite{Coles2018QuantumBeginners} for an implementation of QAOA on the IBMQX4 Quantum Computer.

Since the original QAOA doesn't enable to enforce constraints directly, we propose below two approaches to include the constraints directly in the objective function $f$.

\subsection{Parameters for QAOA}

The choice of parameters $(\beta_k,\gamma_k)$ with $ k \in [\![1 ; p ]\!]$ is subject to discussion. In the original version of QAOA \cite{Farhi2014AAlgorithm}, optimisation of these parameters is introduced, leading to expensive training. We have then chosen not to take into consideration this optimisation part and to set $\beta_k = (1-\frac{k}{p})$ and $\gamma_k = \frac{k}{p}$, corresponding to the linear evolution in the basic adiabatic approach. Further work could investigate the optimal angles for our class of problem. Otherwise, for the MAX-CUT problem, \cite{Crooks2018PerformanceProblem} \cite{Zhou2018QuantumDevices} suggest that optimal angles may be close to linear annealing for certain sets of problems.

\newpage
\section{Constrained Knapsack QAOA approach}
This section introduces a first method that adapts the adiabatic approach proposed in \cite{Coffey2017AdiabaticProblem} to universal gate quantum computers. The Hamiltonian directly includes the constraint through a strong quadratic penalty, hence its optimisation is equivalent to the resolution of the constrained problem.

\subsection{Circuit overview}
We introduce $e = \lfloor \log_2 C_{max} \rfloor + 1$ and we use a register ${R}$ of size $n + e$, where $n$ is the number of time windows, ${R}[t]$ being the t-th element of register $R$. We only require one ancillary qubit $F$ set initially to $\ket{0}$ to facilitate the computation. The Figure \ref{fig:circuit1} below gives an overview of the circuit, illustrating the main steps of the computation.

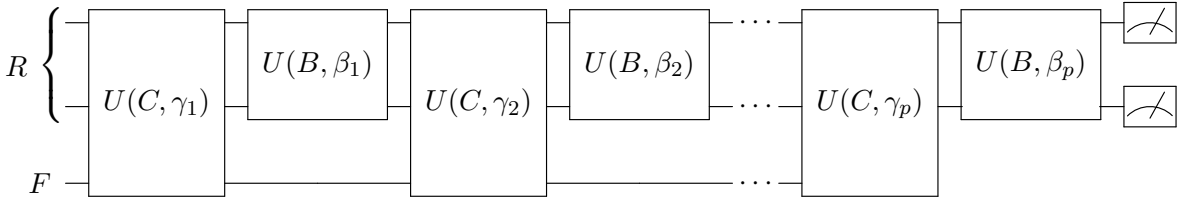
\begin{figure}[h!]
  \centering
  \Qcircuit @C=0.8em @R=1.5em {
  & \multigate{2}{U(C,\gamma_1)} &\multigate{1}{U(B,\beta_1)} & \multigate{2}{U(C,\gamma_2)} &\multigate{1}{U(B,\beta_2)} & \qw &\cdots & &  \multigate{2}{U(C,\gamma_p)} &\multigate{1}{U(B,\beta_p)} & \meter \\
  & \ghost{U(C,\gamma_1)} & \ghost{U(B,\beta_1)} & \ghost{U(C,\gamma_2)} & \ghost{U(B,\beta_2)} & \qw & \cdots & & \ghost{U(C,\gamma_p)} & \ghost{U(C,\beta_p)} & \meter&\\
  \lstick{{F}} & \ghost{U(C,\gamma_1)} & \qw & \ghost{U(C,\gamma_2)} & \qw & \qw & \cdots & & \ghost{U(C,\gamma_p)}&&\\
  \relax\inputgroupv{1}{2}{1em}{1.5em}{{R}}
  }
  \caption{Overview of the circuit of Constrained Knapsack QAOA}
  \label{fig:circuit1}
\end{figure}{}

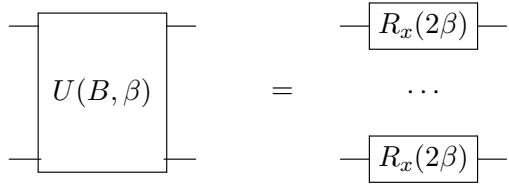
\begin{figure}[h!]
  \centering
  \begin{tabular}{c}
    \Qcircuit @C=1em @R=1.5em{
    \lstick{} & \multigate{2}{{U(B,\beta)}}  & \qw &   &   &   &   &   & \gate{R_x(2 \beta)} & \qw \\
    & & & & & = & & & \cdots\\
    \lstick{} & \ghost{\mathcal{U(B,\beta)}} & \qw &   &   &   &   &   & \gate{R_x(2 \beta)} & \qw \\
    }
  \end{tabular}
  \caption{Details of $U(B,\beta)$ computation}
  \label{fig:Ubeta}
\end{figure}

Knowing that $U(B,\beta) = \exp(-\i\beta B) = \prod_{i=1}^n \exp(-\i\beta \sigma_i^x)$, the Figure \ref{fig:Ubeta} shows how it can be computed using $R_x(2\beta)$ associated to $\exp(-\i\beta \sigma_i^x)$ .

{}

\subsection{Choice of the Hamiltonian}
The Hamiltonian we reintroduce is similar to the one used in \cite{Coffey2017AdiabaticProblem}.

We call $\ket{z} = \ket{z_1z_1...z_n} = {R}[0:n-1]$ and $\ket{b} = \ket{b_0b_1...b_{e}}= R[n:n+e]$. We thus define the objective function of the Hamiltonian as:
\begin{equation}
  f(z,b) = -A\bigg( cost(z) - \sum_{j=0}^{e-1} 2^jb_j - (C_{max} + 1 - 2^d)b_e \bigg)^2 + \sum_{t=1}^n (1-z_t)\lambda_1^t + z_t\lambda_2^t
\end{equation}

If we set $A = \sum_{t=1}^n (\lambda_1^t + \lambda_2^t)$, the maximum of $f$ is the maximum of the problem subject to $cost(z) \leq C_{max}$. It is trivial to see that this problem can be reformulated into an Ising problem, i.e. with the previous notations, it is therefore equivalent to optimise the following function:
\begin{equation}
  f(z,b) = \sum_{0 \leq t \leq n+e} \alpha_t R[t] \hspace{5mm}+ \sum_{0 \leq t_1 < t_2 \leq n+e} \alpha_{t_1,t_2} R[t_1]R[t_2]
\end{equation}

\subsection{Computation of the Hamiltonian}
Since each term commutes, we can compute the Hamiltonian block after block for each term:
\begin{eqnarray}
  U(\gamma,H_f) &=& \exp{(-\i\gamma.f(z,b))} \\
  &=& \exp{\Bigg(-i \gamma \sum_{0 \leq t < n+e} \alpha_t R[t] \hspace{5mm}+ \sum_{0 \leq t_1 < t_2 < n+e} \alpha_{t_1,t_2} R[t_1]R[t_2] \Bigg)}
  \nonumber \\
  &=& \prod_{0 \leq t \leq n+e} \exp{(-i \gamma \alpha_t R[t])} \prod_{0 \leq t_1 < t_2 \leq n+e} \exp{(-i \gamma \alpha_{t_1,t_2} R[t_1]R[t_2])} \nonumber
\end{eqnarray}

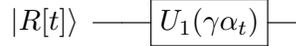
\begin{figure}[H]
  \centering
  \begin{tabular}{c}
    \Qcircuit @C=1em @R=1em{
      & \lstick{\ket{R[t]}} & \qw & \gate{U_1(\gamma \alpha_t)} & \qw &
    }
  \end{tabular}
  \caption{Single term}
  \label{fig:singterm}
\end{figure}

\textit{Single terms.} The terms $\exp{(-i \gamma \alpha_t R[t])}$ can be computed as shown in Figure \ref{fig:singterm}.

\begin{figure}[h!]
  \centering
  \begin{tabular}{c}
    \Qcircuit @C=1em @R=1em{
      & \lstick{\ket{R[t_1]}} & \qw & \ctrl{2}     & \qw                          & \ctrl{2}     & \qw \\
      & \lstick{\ket{R[t_2]}} & \qw & \control \qw & \qw                          & \control \qw & \qw \\
      & \lstick{\ket{0}}      & \qw & \targ        & \gate{U_1(\gamma \alpha_{t_1,t_2})} & \targ        & \qw \\
    }
  \end{tabular}
  \caption{Crossed term}
  \label{fig:crosterm}
\end{figure}
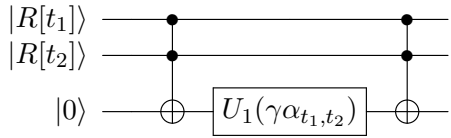

\textit{Crossed terms.} The terms $\exp{(-i \gamma \alpha_{t_1,t_2} R[t_1]R[t_2])}$ can be computed as shown in Figure \ref{fig:crosterm}.

\newpage
\section{Relaxed Knapsack QAOA approach}
\label{sec:relaxed}
As presented in section 5, results of the first constrained approach are unsatisfying for low and middle depth ($p \leq 50$), leading us to developed a novel algorithm. In this second variant, we propose to relax the problem by introducing a penalty for solutions that don't fulfil the constraint. In the case of our battery scheduling problem, that consists in penalising schedules that would exceed the battery health estimate, $C_{max}$.

\subsection{Circuit overview}
We have computed two variants that have different architectures and different lengths but that are based on the same Hamiltonian and have therefore same performances.

\subsubsection{Qubits requirement}

In both variants we require a register ${R}$ of $n$ qubits to store the set of choices and a register $A$ of "ancillary" qubits that we don't need to measure at the end of the circuit. The size and the composition of $A$ differ from one architecture to another as we explain later but we can decompose it in 3 parts: $A_1$ a subregister of size $k_1$ related to cost, $F$ a single flag qubit and $A_2$ of size $k_2$ an additional subregister to compute basic operations.  All the registers are initially set to $\ket{0}$ and at the beginning we turn the qubits of $R$ into the ground state of $H_p$ $\ket{+} = \frac{\ket{0} + \ket{1}}{\sqrt{2}}$.

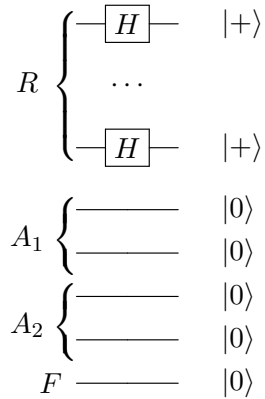
\begin{figure}[h!]
  \centering
  \begin{tabular}{c}
    \Qcircuit @C=1em @R=1.5em {
    \lstick{}  & \gate{H} & \qw & \rstick{\ket{+}} \\
    &\cdots\\
    \lstick{}  & \gate{H} & \qw & \rstick{\ket{+}}
    \inputgroupv{1}{3}{0.8em}{2em}{R}\\
    \lstick{}  & \qw      & \qw & \rstick{\ket{0}} \\
    \lstick{}  & \qw      & \qw & \rstick{\ket{0}}
    \inputgroupv{4}{5}{0.8em}{1em}{A_1}\\
    \lstick{}  & \qw      & \qw & \rstick{\ket{0}} \\
    \lstick{}  & \qw      & \qw & \rstick{\ket{0}}
    \inputgroupv{6}{7}{0.8em}{1em}{A_2}\\
    \lstick{F} & \qw      & \qw & \rstick{\ket{0}} \\
    }
  \end{tabular}
  \vspace{3mm}
  \caption{Initial State}
  \label{fig:initstate}
\end{figure}

\subsubsection{General architecture}

The following figure gives an overview of a circuit of depth p. The architecture is near to the one introduced in the first approach although it needs more ancillary qubits.

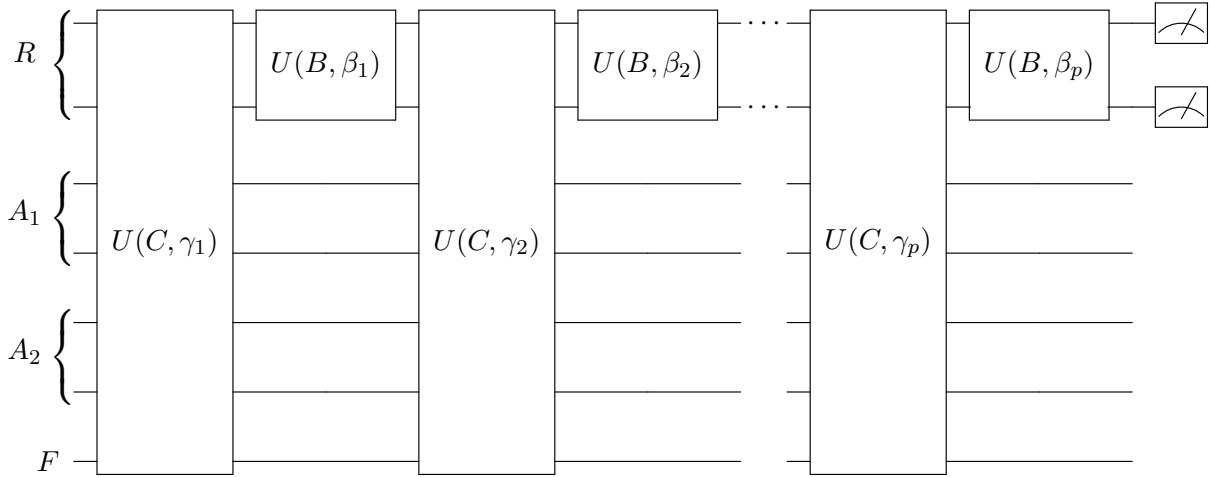
\begin{figure}[h!]
  \centering
  \begin{tabular}{c}
    \Qcircuit @C=0.8em @R=1.5em {
      & \multigate{6}{U(C,\gamma_1)} & \multigate{1}{U(B,\beta_1)} & \multigate{6}{U(C,\gamma_2)} & \multigate{1}{U(B,\beta_2)} & \qw & \cdots &   & \multigate{6}{U(C,\gamma_p)} & \multigate{1}{U(B,\beta_p)} & \qw & \meter \\
      & \ghost{U(C,\gamma_1)}        & \ghost{U(B,\beta_1)}        & \ghost{U(C,\gamma_2)}        & \ghost{U(B,\beta_2)}        & \qw & \cdots &   & \ghost{U(C,\gamma_p)}        & \ghost{U(C,\beta_p)}        & \qw & \meter
    \inputgroupv{1}{2}{.8em}{1em}{R}\\
    & \ghost{U(C,\gamma_1)} & \qw &  \ghost{U(C,\gamma_2)} & \qw & \qw & & & \ghost{U(C,\gamma_p)} & \qw & \qw\\
    & \ghost{U(C,\gamma_1)} & \qw &  \ghost{U(C,\gamma_2)} & \qw & \qw & & &\ghost{U(C,\gamma_p)}
    \inputgroupv{3}{4}{.8em}{1em}{A_1} & \qw & \qw \\
    & \ghost{U(C,\gamma_1)} & \qw &  \ghost{U(C,\gamma_2)} & \qw & \qw & & &\ghost{U(C,\gamma_p)} & \qw & \qw\\
    & \ghost{U(C,\gamma_1)} & \qw &  \ghost{U(C,\gamma_2)} & \qw & \qw & & &\ghost{U(C,\gamma_p)} & \qw & \qw
    \inputgroupv{5}{6}{.8em}{1em}{A_2}\\
    \lstick{F} & \ghost{U(C,\gamma_1)} & \qw &  \ghost{U(C,\gamma_2)} & \qw & \qw & & & \ghost{U(C,\gamma_p)} & \qw & \qw \\
    }
  \end{tabular}
  \caption{Overview of the circuit}
  \label{fig:circuit}
\end{figure}{}

The computation of $U(B,\beta)$ is identical to the previous one.

\subsection{Penalty Hamiltonian}
This second approach consists in penalising linearly too costly set of choices. Given $\ket{z} = \ket{z_1z_2...z_n}$ a set of choices, we introduce $cost(z) = \sum_{t=1}^n (1-z_t)c_1^t + z_tc_2^t$ the number of cycles associated with $\ket{z}$. We define the return function $return(z)$ and the penalty function $penalty(z)$ as follows:
\begin{equation}
  return(z) =  \sum_{t=1}^n return_t(z) \hspace{5mm} with \hspace{2mm} return_t(z) \equiv (1-z_t)\lambda_1^t + z_t\lambda_2^t
\end{equation}
\begin{center}
  \[
    penalty(z) = \hspace{4mm}
    \begin{cases}
      0 \hspace{33mm} if \hspace{2mm}cost(z) < C_{max}                                                             \\
      -\alpha(cost(z) - C_{max}) \hspace{2mm} if \hspace{2mm}cost(z) \geq C_{max}, \alpha > 0\hspace{2mm} constant \\
    \end{cases}
  \]
\end{center}

And we define the objective function:
\begin{equation}
  f(z) = return(z) + penalty(z)
\end{equation}

It is a classical penalty approach with a penalty mounted the absolute value of the error.

\subsection{Computation of the return part}
We detail here the realisation of the circuit.
\begin{eqnarray}
  U(C,\gamma)\ket{z} & =&  e^{-i \gamma f(z)}\ket{z} \\
  & = & e^{-i \gamma .penalty(z)}e^{-i \gamma .return(z)}\ket{z} \nonumber
\end{eqnarray}

We compute first the return part:
\begin{eqnarray}
  e^{-i \gamma .return(z)}\ket{z} &=& \prod_{t=1}^n e^{-i \gamma return_t(z)}\ket{z} \\
  &= &e^{i \theta} \bigotimes_{t=1}^n e^{-i \gamma z_t(\lambda_2^t - \lambda_1^t)}\ket{z_t} \nonumber\\
  \text{with} ~ \hspace{2mm} \theta &= &\sum_{t=1}^n \lambda_1^t \hspace{3mm} \text{constant} \nonumber
\end{eqnarray}

As the phase is $\theta$ is independent from $z$, it doesn't change the final distribution and we can ignore it in the computation. We compute in parallel the terms $e^{-i \gamma (\lambda_2^t - \lambda_1^t)z_t}\ket{z_t}$ as follows.

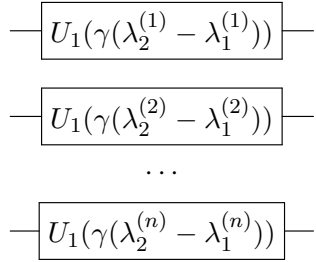
\begin{figure}[h!]
  \centering
  \begin{tabular}{c}
    \Qcircuit @C=1em @R=1em {
      & \gate{U_1(\gamma(\lambda_2^{(1)}-\lambda_1^{(1)}))} & \qw \\
      & \gate{U_1(\gamma(\lambda_2^{(2)}-\lambda_1^{(2)}))} & \qw \\
    &\cdots\\
      & \gate{U_1(\gamma(\lambda_2^{(n)}-\lambda_1^{(n)}))} & \qw \\
    }
  \end{tabular}
  \caption{Return Hamiltonian}
  \label{fig:retham}
\end{figure}{}

\subsection{Computation of the penalty part}
The penalty Hamiltonian is harder to compute for this relaxed approach. We present below two implementations to compute $e^{-i \gamma .return(z)}$. In both cases, we decompose its computation in 4 subroutines:
\begin{enumerate}
  \item \textit{Cost calculation.} We compute the entangled states $\ket{z}\otimes\ket{cost(z)}$.
  \item \textit{Test constraint.} We test the condition $cost(z) < C_{max}$ and we set the flag $F$ to $1$ if the inequality is unsatisfied. We obtain then the entangled state $\ket{z}\otimes\ket{cost(z)}\otimes\ket{\big(C_{max}\leq cost(z)\big)}$.
  \item \textit{Penalty dephasing.} We rotate of $e^{i \gamma \alpha (cost(z) - C_{max})}$ under control of the flag qubit.
  \item \textit{Reinitialisation.} We undo operations 2 and 1 to reset the ancillary qubits.
\end{enumerate}

\subsubsection{Subroutine 1: Cost calculation}
Given a set of choices $\ket{z_1...z_n}$, subroutine 1 stores the associated cost in $\Tilde{A_1}$, a subregister of register $A_1$. As stated in Equation \ref{eq:maxhealth}, there is a constant $d$ such that $c_i^t \leq d, \hspace{2mm} \forall t \in [\![1,n]\!]$. We present two ways to compute the cost, the second one being more accurate, but harder to compute and needs a larger register entanglement as well.

\textit{Variant 1.} In this variant, $A_1 = \Tilde{A_1}$, i.e. we store $\ket{cost(z)}$ the total cost in $A_1$. It is set to $\ket{0}^{k_1}$ at the beginning, and then the cost is incremented successively for each time window $t$. The details of the subroutine $add(x)$ in Figure \ref{fig:subadd}, that adds $x$ to the register $A_1$, can be found in Appendix A.

\begin{figure}[h!]
  \centering
  \begin{tabular}{c}
    \Qcircuit @C=1em @R=1em{
    \lstick{\ket{z}} & \gate{add(x)} & \rstick{\ket{z+x}} \qw \\
    }
  \end{tabular}
  \caption{Subroutine $add$, see Appendix A.}
  \label{fig:subadd}
\end{figure}

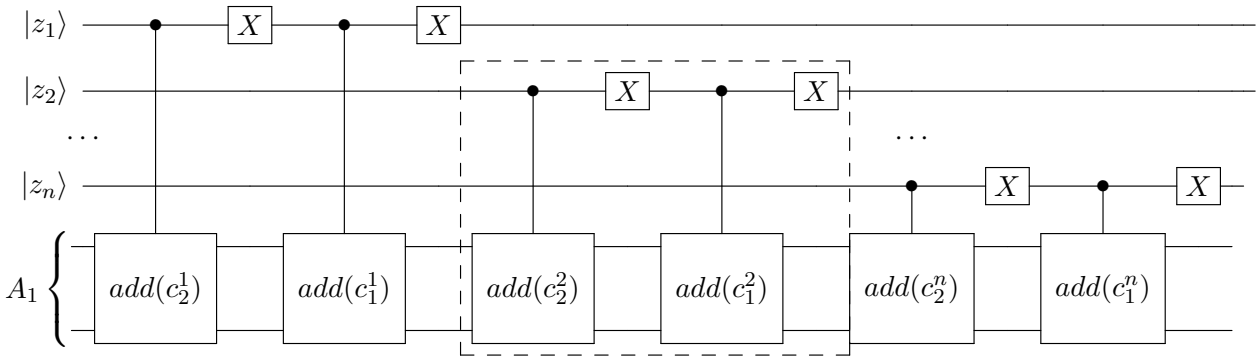
\begin{figure}[h!]
  \centering
  \begin{tabular}{c}
    \Qcircuit @C=0.4em @R=1em {
      & \lstick{\ket{z_1}} & \ctrl{4} & \gate{X} & \ctrl{4} & \gate{X} & \qw      & \qw      & \qw      & \qw      & \qw & \qw      & \qw      & \qw      & \qw      & \qw & \qw \\
      & \lstick{\ket{z_2}} & \qw      & \qw      & \qw      & \qw      & \ctrl{3} & \gate{X} & \ctrl{3} & \gate{X} & \qw & \qw      & \qw      & \qw      & \qw      & \qw & \qw \\
    & \cdots & & & & & & & & &\cdots\\
      & \lstick{\ket{z_n}} & \qw      & \qw      & \qw      & \qw  & \qw & \qw  & \qw & \qw & \ctrl{1} & \gate{X} & \ctrl{1} & \gate{X} & \qw & \qw \\
    &\lstick{} \qw &\multigate{2}{add(c_2^1)} &  \qw & \multigate{2}{add(c_1^1)} & \qw &\multigate{2}{add(c_2^2)} & \qw & \multigate{2}{add(c_1^2)} & \qw  & \multigate{2}{add(c_2^n)} & \qw & \multigate{2}{add(c_1^n)} & \qw & \qw \\
    & \\
    &\lstick{} \qw & \ghost{add(c_1^1)} & \qw & \ghost{add(c_1^2)} & \qw & \ghost{add(c_2^1)} & \qw & \ghost{add(c_2^2)} & \qw  & \ghost{add(c_1^n)} & \qw & \ghost{add(c_2^n)} & \qw & \qw \\
    \relax\inputgroupv{5}{7}{1em}{1.5em}{A_1} \\
    \relax\gategroup{2}{7}{7}{10}{0.8em}{--}
    }
  \end{tabular}
  \caption{Cost calculation. \textit{Variant 1.}}
  \label{fig:sub1tech1}
\end{figure}

The cost can then be calculated and stored as shown in Figure \ref{fig:sub1tech1}: if $\ket{z_2} = \ket{0}$, $A_1$ is incremented of $c_1^2$, and if $\ket{z_2} = \ket{1}$ $A_1$ is incremented of $c_2^2$. We repeat thus the operation for every $\ket{z_t}$ and at the end $A_1$ contains $\ket{cost(z)}$.

\textit{Variant 2.} We assume here that $A_1$ has $n$ subregisters $A_1^i, \forall i \in [\![1,n]\!]$ and we store eventually the total cost in $A_1^1 = \Tilde{A_1}$. We suppose we dispose of another subroutine $\barr $ such that:

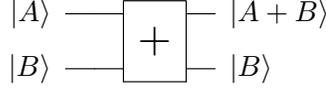
\begin{figure}[h!]
  \centering
  \begin{tabular}{c}
    \Qcircuit @C=1em @R=1em{
      & \lstick{\ket{A}} & \qw & \multigate{1}{\barr} & \rstick{\ket{A+B}} \qw \\
      & \lstick{\ket{B}} & \qw & \ghost{\barr}        & \rstick{\ket{B}} \qw   \\
    }
  \end{tabular}
  \caption{Subroutine $\barr$}
  \label{fig:sub}
\end{figure}

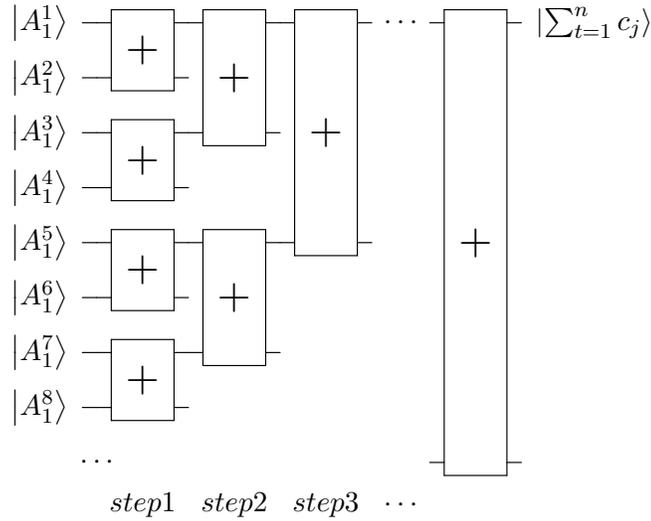
\begin{figure}[h!]
  \centering
  \begin{tabular}{c}
    \Qcircuit @C=0.5em @R=1em {
    \lstick{\ket{A_1^1}} & \qw    & \multigate{1}{\barr} & \qw & \multigate{2}{\barr} & \qw & \multigate{4}{\barr} & \qw &   & \cdots &   &   & \multigate{8}{\barr} & \rstick{\ket{\sum_{t=1}^n c_j}} \qw \\
    \lstick{\ket{A_1^2}} & \qw & \ghost{\barr} &  \qw\\
    \lstick{\ket{A_1^3}} & \qw &\multigate{1}{\barr} &  \qw & \ghost{\barr} & \qw\\
    \lstick{\ket{A_1^4}} & \qw & \ghost{\barr} &  \qw\\
    \lstick{\ket{A_1^5}} & \qw & \multigate{1}{\barr} &  \qw & \multigate{2}{\barr} & \qw & \ghost{\barr} & \qw\\
    \lstick{\ket{A_1^6}} & \qw & \ghost{\barr} & \qw\\
    \lstick{\ket{A_1^7}} & \qw & \multigate{1}{\barr} & \qw & \ghost{\barr} & \qw\\
    \lstick{\ket{A_1^8}} & \qw & \ghost{\barr} & \qw \\
                         & \cdots &                      &     &                      &     &                      &     &   &        &   &   & \ghost{\barr}        & \qw                                 \\
    & & step 1 & & step 2 & & step 3 & & & \cdots \\
    }
  \end{tabular}
  \caption{Cost calculation. \textit{Variant 2.}}
  \label{fig:sub1tech2}
\end{figure}

To achieve the computation, illustrated on Figure \ref{fig:sub1tech2}, we first store in each subregister $A_1^i$ the cost associated with choice $z_i$: $c^i_1 \rightarrow \ket{A_1^i}$ if $\ket{z_i} = 0$, and $c^i_2 \rightarrow \ket{A_1^i}$ if $\ket{z_i} = 1$. Then we use a technique inspired by parallelism to sum all terms.

{\textit{Size of the registers.}}
For both technique 1 and 2, we have to impose that every register is large enough to store the number it has to store, otherwise it creates an overflow. For \textit{variant 1}, we must impose $A_1$ big enough to store the largest cost. While every cost are less than $d$, we notice $cost(z) \leq dn, \forall z$, where $d = \max c_i^t$. Thus by imposing $A_1$ of size $k_1 = \lfloor \log_2 dn \rfloor + 1 = O(\log_2 n)$, we can ensure there is no overflow.

The size of the registers for \textit{variant 2} is harder to determine. We give the details in Appendix B, but we sketch the idea here. We note that:
\begin{itemize}
  \item $\ket{A_1^2}, \ket{A_1^4}, \ket{A_1^6}...$ have to store 1 cost and must thus be of size $\lfloor \log_2(d) \rfloor +1$.
  \item $\ket{A_1^3}, \ket{A_1^7}, \ket{A_1^{11}}...$ have to store the sum of 2 costs and must thus be of size $\lfloor \log_2(2d) \rfloor +1 = \lfloor \log_2(d) \rfloor + 2 $.
  \item $\ket{A_1^5}, \ket{A_1^{13}}, ...$ have to store the sum of 4 costs and must thus be of size $\lfloor \log_2(4d) \rfloor +1 = \lfloor \log_2(d) \rfloor + 3 $.
  \item ...
\end{itemize}
And the largest subregister $A_1^1$ has to be of size $O(\log_2 n)$ as well.

\subsubsection{Subroutine 2: Constraint testing}
Subroutine 2 tests the condition $cost(z) < C_{max}$ and set $F$ to $1$ if not. At that stage, we have already computed the cost, stored in $\Tilde{A_1}$ a subregister of $A_1$, that can be expressed in binary writing: $cost(z) = \sum_{j=0}^{k_1-1} 2^{j}\Tilde{A_1}[j]$.

Let's first assume $C_{max}=2^c$ where $c$ is an integer. $cost(z) < C_{max} \iff \bigcap^{k_1-1}_{j=c} \big(\Tilde{A_1}[j] = 0\big)$. This multiple condition could be tested with a $(k_1 - c)$-NOT gate. The circuit presented in Figure \ref{fig:sub2} set $F$ to $1$ if and only if $\bigcap^{k_1-1}_{t=c} \big(\Tilde{A_1}[t] = 0\big) \iff \bigcap^{k_1-1}_{t=c} \big(\neg \Tilde{A_1}[t] = 1\big) $. It is equivalent to a multiple C-NOT gate. If $c \geq k_1$ the inequality is always satisfied.

\begin{figure}[h!]
  \centering
  \begin{tabular}{c}
    \Qcircuit @C=1em @R=1.5em {
    &\lstick{\Tilde{A_1}[0]} & \qw & \qw & \qw & \qw &\qw &\qw\\
    &\cdots & \\
    &\lstick{\Tilde{A_1}[c-1]} & \qw & \qw  & \qw & \qw & \qw  & \qw\\
    &\lstick{\Tilde{A_1}[c]} & \gate{X} & \ctrl{5}  & \qw & \qw & \qw  & \qw\\
    &\lstick{\Tilde{A_1}[c+1]} & \gate{X} & \control \qw & \qw & \qw & \qw & \qw\\
    &\lstick{\Tilde{A_1}[c+2]} & \gate{X} & \qw & \ctrl{4} & \qw & \qw & \qw\\
    &\cdots & & & \cdots\\
    &\lstick{\Tilde{A_1}[k_1 - 1]} & \gate{X} & \qw & \qw & \qw & \ctrl{5} & \qw\\
    &\lstick{A_2[0] = \ket{0}} & \qw & \targ & \control \qw & \qw & \qw & \qw\\
    &\lstick{A_2[1] = \ket{0}} \qw & \qw & \qw & \qw & \targ & \qw & \qw & \qw\\
    & & \cdots\\
    & \lstick{A_2[(k_1 - c - 3)] = \ket{0}} & \qw & \qw & \qw & \qw & \control \qw & \qw\\
      & \lstick{F} & \qw & \qw & \qw & \qw & \targ & \qw \\
    }
  \end{tabular}
  \caption{Test constraint circuit.}
  \label{fig:sub2}
\end{figure}
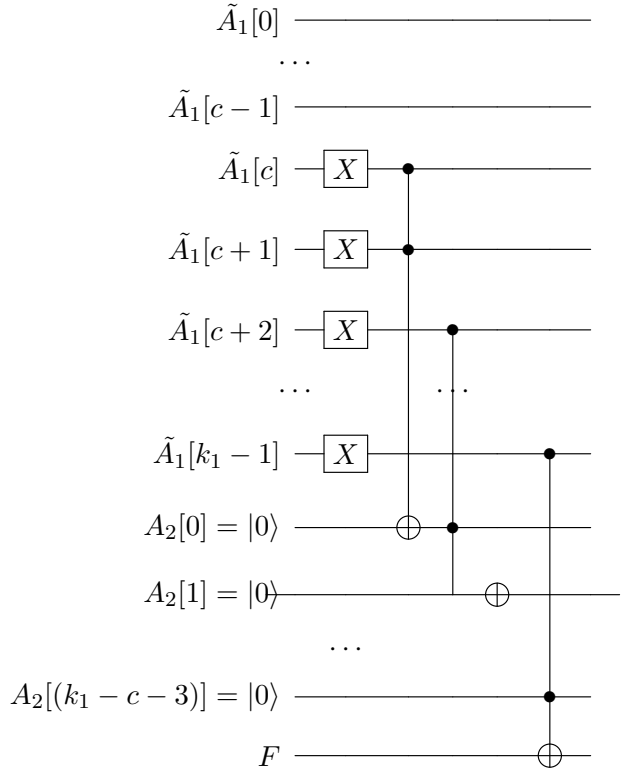{}

Suppose we are now in the general case, where $C_{max}$ can be any integer, not only powers of 2. We simply add a constant $w$ to be in the previous case: $cost(z) < C_{max} \iff cost(z) + w < C_{max} + w = 2^c$, see Appendix A for the subroutine to do and undo additions.

\subsubsection{Subroutine 3: Penalty dephasing}
Given the previous trick, let still assume that $C_{max} = 2^c$, according that $(C_{max}+w) - (cost(z)+w) = C_{max} - cost(z)$. We have:
\begin{equation}
  \alpha(cost(z) - C_{max}) = \sum_{j=0}^{k_1-1}2^j\alpha A_1[j] - 2^c\alpha
\end{equation}

Therefore, the subroutine 3 can be computed with the circuit below.

\begin{figure}[h!]
  \centering
  \begin{tabular}{c}
    \Qcircuit @C=1em @R=1.5em {
    \lstick{\Tilde{A_1}[0]}       & \gate{U_1(2^0\alpha \gamma)} & \qw                          & \qw                                & \qw                             & \qw \\
    \lstick{\Tilde{A_1}[1]}       & \qw                          & \gate{U_1(2^1\alpha \gamma)} & \qw                                & \qw                             & \qw \\
    & & & \cdots \\
    \lstick{\Tilde{A_1}[k_1 - 1]} & \qw                          & \qw                          & \gate{U_1(2^{k_1-1}\alpha \gamma)} & \qw                             & \qw \\
    \lstick{F}                    & \ctrl{-4}                    & \ctrl{-3}                    & \ctrl{-1}                          & \gate{U_1(- 2^c \alpha \gamma)} & \qw \\
    }
  \end{tabular}
  \caption{Penalty dephasing.}
  \label{fig:sub3}
\end{figure}
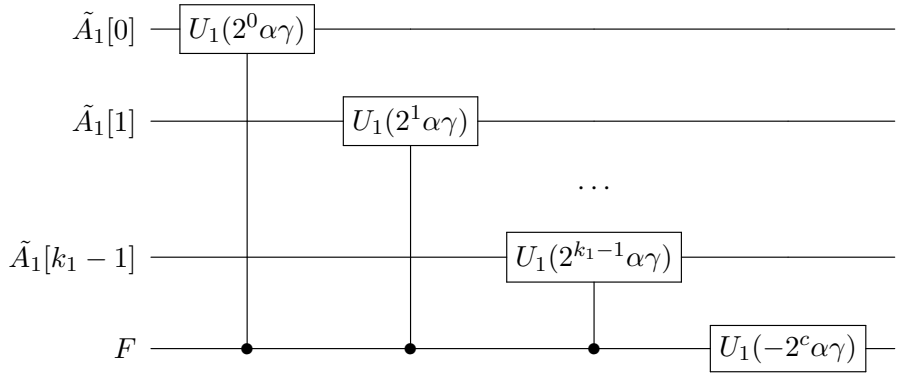{}

\subsubsection{Subroutine 4: Reinitialisation}
At the end of the penalty computation, we want to set the ancillary qubits back in their initial state where they are all set to $\ket{0}$. Subroutines 1 and 2 are composed only with X-gates and CNOT-gates, and both are involutive (they are their own reciprocal). We thus compute the circuit in reverse direction to reset the ancillary qubits.

\subsection{Discussion and complexity}

\subsubsection{Complexity and required qubits of the different subroutines.}
The computation of the return part is computed in time $O(1)$ and needs no ancillary qubits. It is the same to compute the block $U(B,\beta)$. Subroutine 2 and subroutine 3 operate in time $O(k_1) = O(\log_2 n)$ and requires at most $k_1$ ancillary qubits. We see below that the global complexity and performances of the circuit is highly dependent of the depth of subroutine 1.

\subsubsection{Complexity of subroutine 1}

\textit{Variant 1.} Subroutine $add$ is computed in time $\log_2(d)size(A_1) = \log_2(d) k_1 = O(\log_2(n))$, as $d$ is a constant (see appendix A for the depth of $ADD$). Since circuit 1 achieves $2n$ blocks $ADD$, \textit{technique 1} requires a circuit of size $O(p n \log n)$. Every operation can be computed with $k_1$ ancillary qubits.

\textit{Variant 2.} If register $A$ of size $a$ and $B$ of size $b$, it takes times $ab$ to compute $\ket{A} \barr \ket{B}$ and needs $a$ ancillary qubits (see appendix A). While the largest register $\ket{A_1^1}$ is of size $\lfloor \log_2(dn) \rfloor + 1 = O(\log_2(n))$, it takes time at most $O(log_2(n)^2)$ to compute each subroutine. In one step, all operations are computed in parallel, with $O(\log_2(n))$ steps (see Appendix B). The subroutine 2 takes therefore a shortest time: $O(\log_2(n)^3)$. However it needs more qubits than \textit{variant 1}: as we compute the operations in parallel during one step, each of them needs different ancillary qubits. We use the fact that all the subregisters are of size $\leq \lfloor \log_2(dn) \rfloor + 1 = O(\log_2(n))$. Step 1 executes $n/2$ operations in parallel. We impose then to have $\frac{n}{2}.\log_2(n) = O(n\log_2(n))$ ancillary qubits.

\subsubsection{Overall complexity and qubits need}
The following tables summarises the performance of each variant.

\begin{table}[h!]
  \centering
  \begin{tabular}{|c|c|c|}
    \hline
    Variant & Depth              & Number of ancillary qubits \\
    \hline
    1       & $O(pn \log_2 n)$   & $O(n)$                     \\
    \hline
    2       & $O(p(\log_2 n)^3)$ & $O(n\log_2n)$              \\
    \hline
  \end{tabular}
  \caption{Performances of each technique}
  \label{tab:perform}
\end{table}

The variant 2 is thus especially interesting since it achieves a $p$-polylogarithmic complexity in $n$. As mentioned above, influence of depth $p$ of QAOA on accuracy of the result is not yet known. However, it is interesting to point out that a classical $\epsilon$-approximation algorithm for the Knapsack problem runs in time $O(\frac{n^3}{\epsilon})$ \cite{Gupta2005ApproximationsProgramming}.

\section{Simulation and Results}

\subsection{Implementation and simulation parameters}
Both algorithms have been implemented with IBM's Qiskit library, and experimented on the QASM simulator \cite{Qiskit}, for different ranges of $n$ and $p$. For each choice $n,p$, we have run the algorithm $1000$ times with randomly chosen values for $\lambda_{1,2}^t$ and $c_{1,2}^t$. $\lambda_1^t$ have been chosen randomly in the range $[\![0,5]\!]$, $\lambda_2^t$ in the range $[\![0,3]\!]$, $c_1^t$ in $[\![0,2]\!]$ and $c_2^t$ in $[\![0,1]\!]$. And we have set $C_{max} = n$ so that a significant number of set of choices are over this value. In the following, we call ratio the value of the output of the algorithm over the optimal value \textit{for the chosen target function}.

\subsection{Results for the constrained approach}
The following chart illustrates the results obtained on the simulator. The algorithm outputs in average a $0.75$ approximation of the optimum. The results are stable while $p$ increases for these range of middle depth under $p = 50$. It is perhaps a consequence of the heavy weight of the constraint compared to the return part in the first Hamiltonian introduced.

\begin{figure}[h!]
  \centering
  \includegraphics[scale = 0.6]{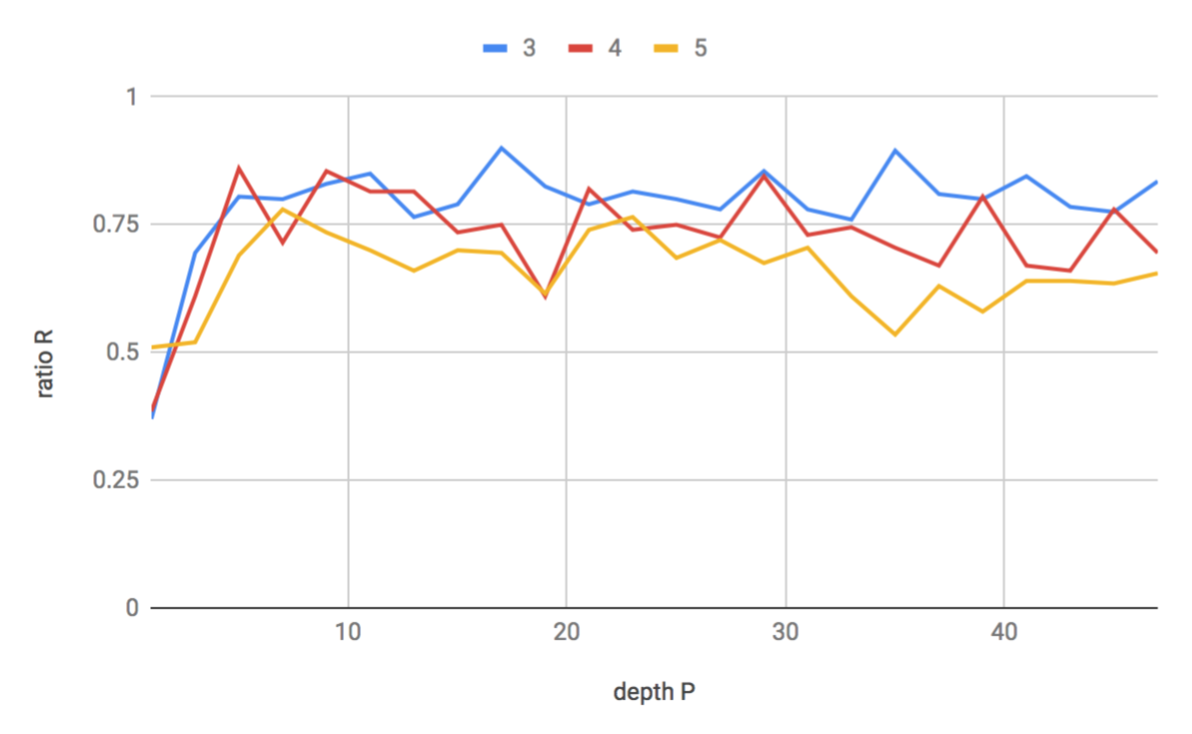}
  \caption{Ratio vs P for each depth n}
  \label{fig:rationPn}
\end{figure}

\subsection{Results for the relaxed approach}

\subsubsection{Without Penalty}
We present here the output without any penalty in the target function(i.e. with only the return part of the circuit). The circuit is much simpler, allowing us to run simulations for bigger circuit, with $n \in \left[1;11\right]$ and $p \in \left[2;8\right]$, see in Appendix C Table \ref{fig:resultswithout}.

Figure \ref{tab:results} shows how the ratio evolves depending on $P$ the depth of the circuit. We observe that the ratio is relatively stable for a given depth depending on $n$. We check also that the precision increases with the depth of the circuit. The figure \ref{fig:fig2} reveals an exponential improvement with $p$. The precision is unfortunately not good enough to test it on bigger input.

\begin{figure}[H]
  \centering
  \includegraphics[scale = 0.6]{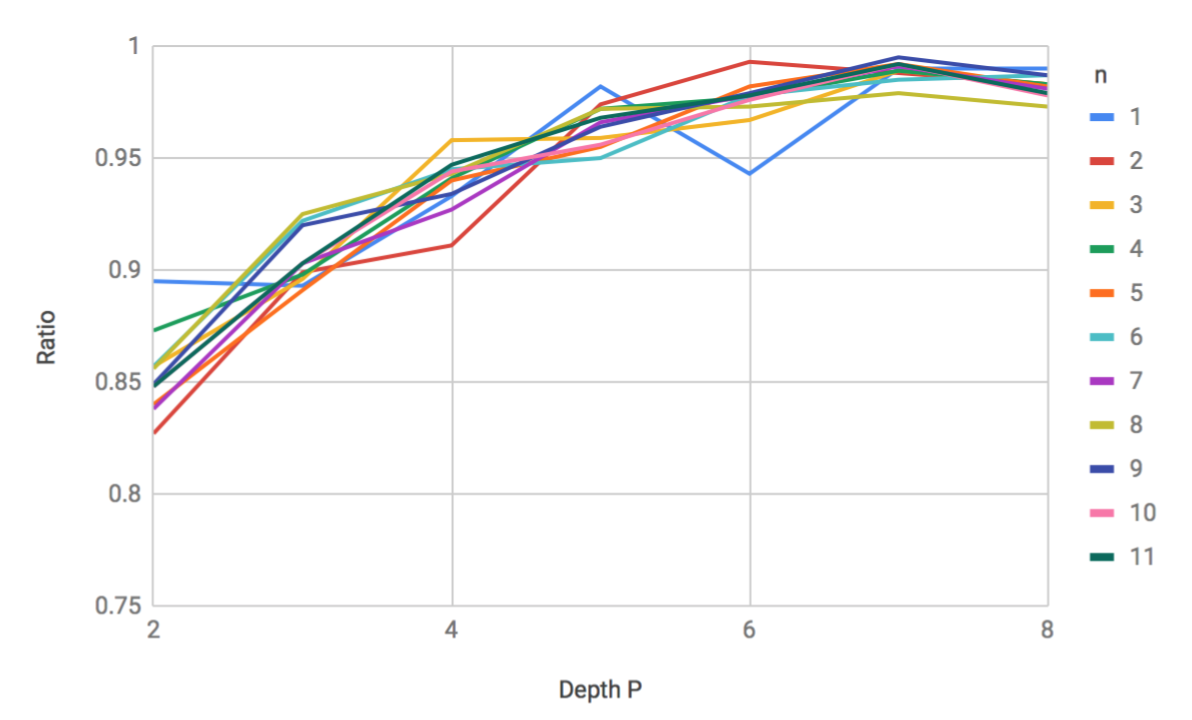}
  \caption{Relaxed Knapsack QAOA without penalty: Ratio vs n, for each depth p (color encoded)}
  \label{tab:results}
\end{figure}
\begin{figure}[H]
  \centering
  \includegraphics[scale = 0.6]{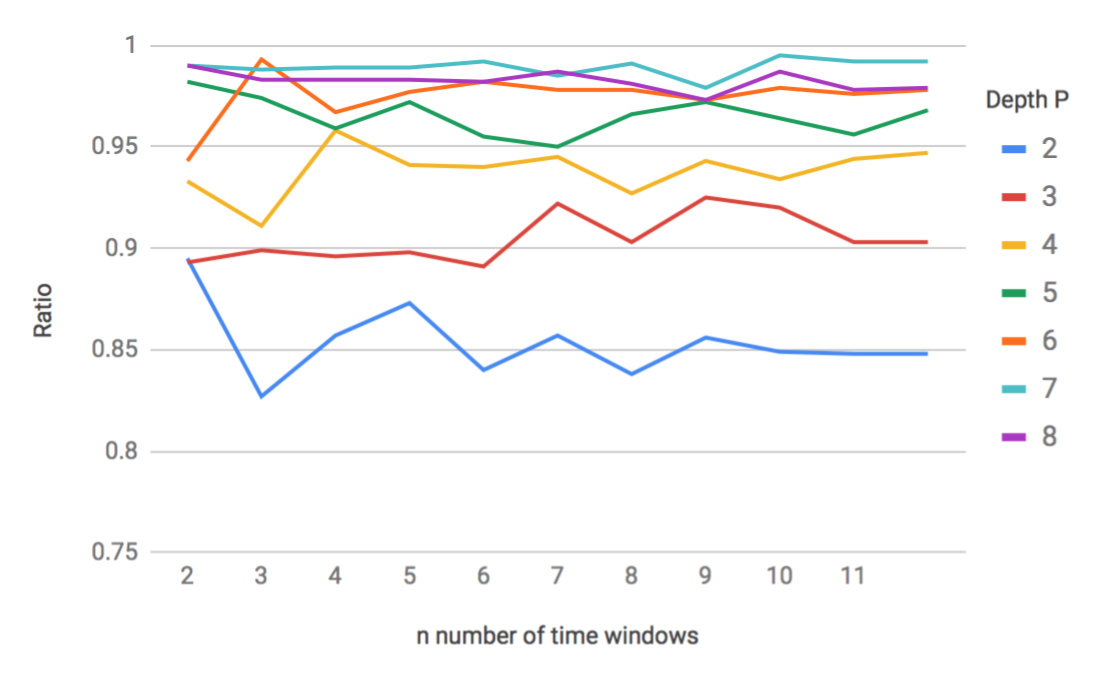}
  \caption{Relaxed Knapsack QAOA without penalty: Ratio vs P, for each number of time windows $n$ (color encoded)}
  \label{fig:fig2}
\end{figure}

\subsubsection{With penalty}
Due to the higher complexity, the simulations launched with the penalty circuit were carried out only for $n \in \left[2;7\right]$ but with $p \in \left[3;12\right]$. We have set here $\alpha = 1$. This value is arbitrary and one can argue that this value should be homogeneous to a return/cost. However, this general investigation enables to test the general impact of the penalty. A further work could investigate the impact of this value on the performance of  the algorithm. 

We observe globally the same pattern as in the non penalised experiments: the performance obviously improves with $P$. It is hard to determine if the performance is stable with that much data. It seems to sink for $n=8$ but we don't know how it behaves for bigger $n$. The exponential is still observable although it is not as clear as in the previous case.

\begin{figure}[H]
  \centering
  \includegraphics[scale = 0.6]{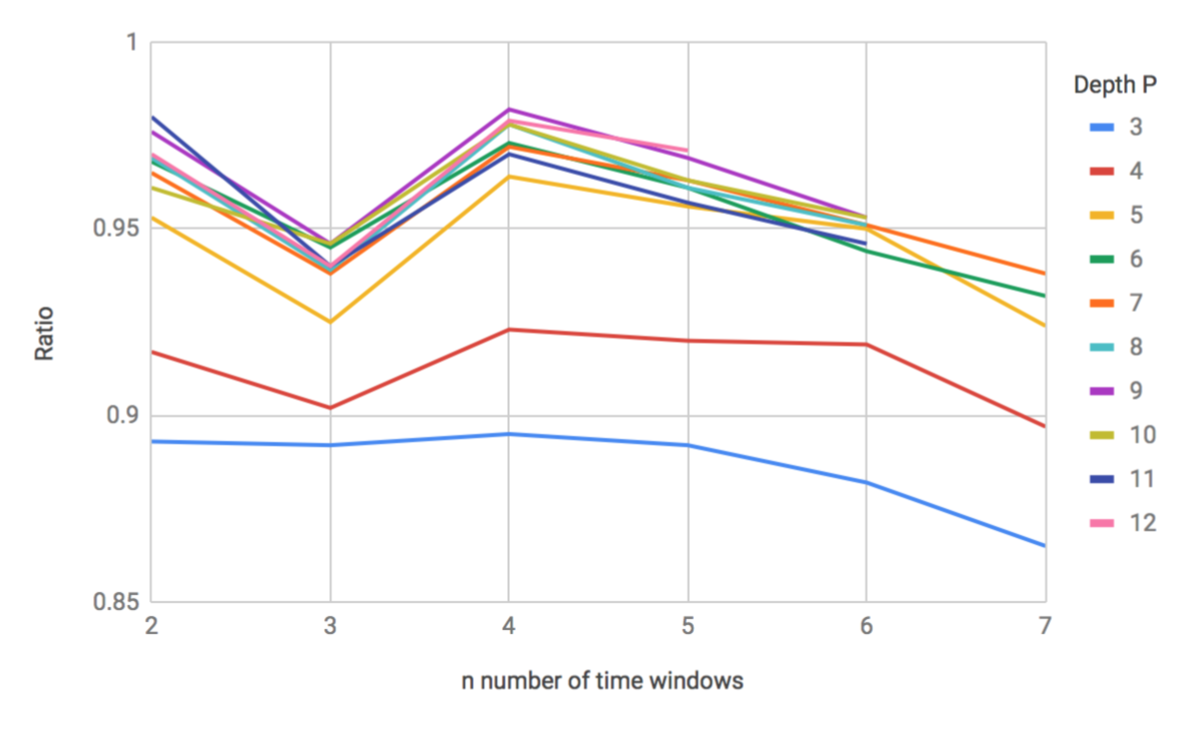}
  \caption{Relaxed Knapsack QAOA with penalty: Ratio vs n, for each depth p (color encoded)}
  \label{fig:Table_penalty_ratio_vs_n}
\end{figure}

\begin{figure}[H]
  \centering
  \includegraphics[scale = 0.6]{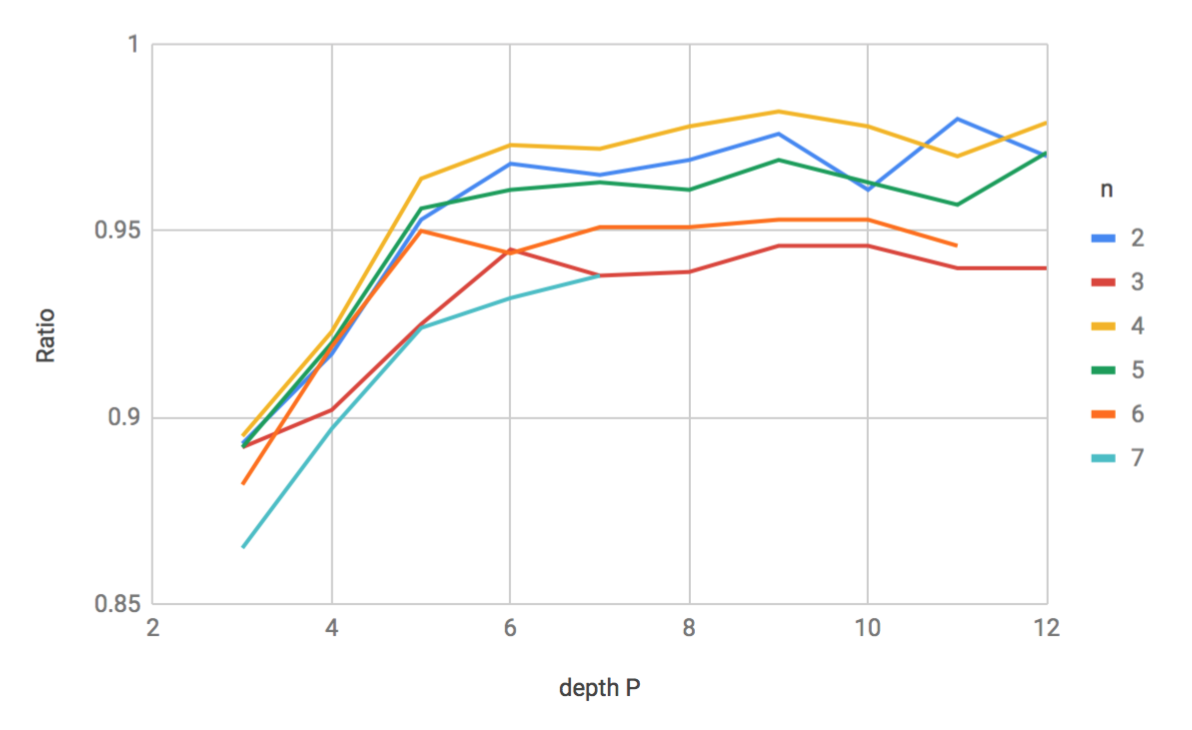}
  \caption{Relaxed Knapsack QAOA with penalty: Ratio vs P, for each number of time steps $n$ (color encoded)}
  \label{fig:Table_penalty_ratio_vs_P}
\end{figure}

\newpage
\section{Conclusion}
We have developed two variants of QAOA to solve the problem of battery scheduling optimisation, modelled here as a Knapsack Problem. This investigation is full of insights about how to tackle a constrained problem with the QAOA. The disappointing results of the 'constrained' approach bring to light that a quadratic penalty may not work properly with shallow depth circuits. We suspect the constraint to be too 'heavy' in the objective function compared to the return part, but it would be a interesting work to investigate further. Nevertheless, the linear penalty has appeared more successful. This is a good illustration that QAOA could work with complex \textit{non-Ising} function with shallow depth circuits. Furthermore, it offers a technique to take linear constraints into account to tackle generic optimisation problems. An evaluation of the influence of $\alpha$, the penalty coefficient, on the behaviour of the algorithm  would be as well of great interest.

It should be pointed out that neither the model nor the experiment take into account the physical issues of current real devices. Hence, if the experimental results cannot be interpreted as a proof of use, they are encouraging and constitute a step further towards coming applications of quantum computing to industrial problems.

\newpage
\section*{Appendices}
\addcontentsline{toc}{section}{Appendices}
\subsection*{Appendix A. Addition subroutine}
\addcontentsline{toc}{subsection}{Appendix A. Addition subroutine}
Different papers have been published to describe the computation of an addition on a quantum computer \cite{Vedral1996QuantumOperations} \cite{Haner2016FactoringMultiplication}. Taken into account are the length of the circuit and the number of ancillary qubits needed for this purpose. In our case, to achieve an operation on a $n$-digit number, we compute a basic addition circuit of length $n$ with $n$ "clean" ancillary qubits.

\subsubsection*{ADD block.}
Suppose we have $\ket{z} = \ket{z_0z_1...z_n}$ a binary number. We want to compute $\ket{z+2^k}$ under control of qubit $\ket{C}$ with a set of $n$ "clean" ancillary qubits all set initially to $\ket{0}$. The circuit below performs this operation in linear time. The ancillary qubits enable to propagate the carry in the addition.

Once we can add a number $2^k$, it is easy to add an arbitrary constant by adding its binary decomposition.

\begin{figure}[h!]
  \centering
  \begin{tabular}{c}
    \Qcircuit @C=1em @R=1em{
      & \lstick{\ket{C}}       & \qw & \ctrl{8}     & \qw          & \qw          & \qw       & \qw          & \qw       & \qw          & \qw       & \ctrl{8}     & \qw \\
      & \lstick{\ket{z_0}}     & \qw & \qw          & \qw          & \qw          & \qw       & \qw          & \qw       & \qw          & \qw       & \qw          & \qw \\
    ...\\
      & \lstick{\ket{z_k}}     & \qw & \control \qw & \qw          & \qw          & \qw       & \qw          & \qw       & \qw          & \targ     & \control \qw & \qw \\
      & \lstick{\ket{z_{k+1}}} & \qw & \qw          & \ctrl{5}     & \qw          & \qw       & \qw          & \qw       & \ctrl{5}     & \qw       & \qw          & \qw \\
    ... \\
      & \lstick{\ket{z_{n-1}}} & \qw & \qw          & \qw          & \ctrl{6}     & \qw       & \ctrl{6}     & \targ     & \qw          & \qw       & \qw          & \qw \\
      & \lstick{\ket{z_n}}     & \qw & \qw          & \qw          & \qw          & \targ     & \qw          & \qw       & \qw          & \qw       & \qw          & \qw \\
      & \lstick{\ket{0}}       & \qw & \targ        & \control \qw & \qw          & \qw       & \qw          & \qw       & \control \qw & \ctrl{-5} & \targ        & \qw \\
      & \lstick{\ket{0}}       & \qw & \qw          & \targ        & \qw          & \qw       & \qw          & \qw       & \targ        & \qw       & \qw          & \qw \\
    ...\\
      & \lstick{\ket{0}}       & \qw & \qw          & \qw          & \control \qw & \qw       & \control \qw & \ctrl{-5} & \qw          & \qw       & \qw          & \qw \\
      & \lstick{\ket{0}}       & \qw & \qw          & \qw          & \targ        & \ctrl{-5} & \targ        & \qw       & \qw          & \qw       & \qw          & \qw \\
    }
  \end{tabular}
  \caption{Crossed term}
  \label{fig:haha}
\end{figure}
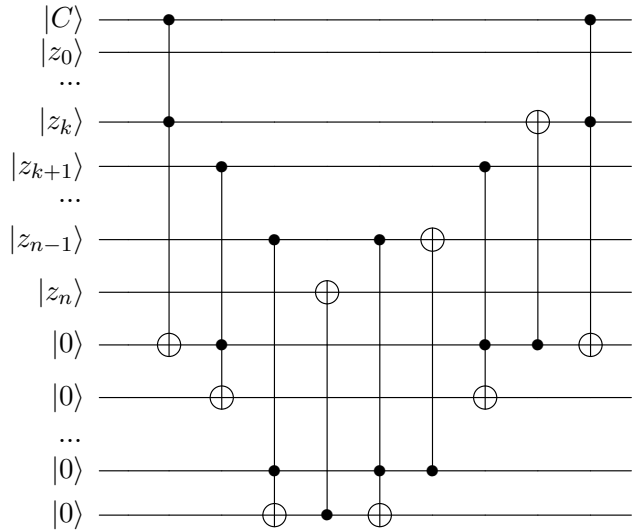
The circuit is of size $O(n)$, and requires $n$ \textit{clean} ancillary qubits, which are set set at $\ket{0}$ at the beginning and at the end.

\subsubsection*{$\barr$ block.}
Suppose now we have $\ket{A} = \ket{a_1a_2...a_p}$ and $\ket{B} = \ket{b_1b_2...b_q}$ and we want to execute the following block:

\begin{figure}[h!]
  \centering
  \begin{tabular}{c}
    \Qcircuit @C=1em @R=1em{
      & \lstick{\ket{A}} & \qw & \multigate{1}{\barr} & \rstick{\ket{A+B}} \qw \\
      & \lstick{\ket{B}} & \qw & \ghost{\barr}        & \rstick{\ket{B}} \qw   \\
    }
  \end{tabular}
  \caption{Subroutine $\barr$}
  \label{fig:sub+}
\end{figure}

The following circuit uses the $ADD$ block exposed above to compute the $\barr$ block.

\begin{figure}[h!]
  \centering
  \begin{tabular}{c}
    \Qcircuit @C=1em @R=1em{
      & \lstick{\ket{A}}   & \qw & \gate{ADD(2^0)} & \gate{ADD(2^1)} & \qw & \cdots &   & \gate{ADD(2^q)} & \qw \\
      & \lstick{\ket{b_1}} & \qw & \ctrl{-1}       & \qw             & \qw &        &   & \qw             & \qw \\
      & \lstick{\ket{b_2}} & \qw & \qw             & \ctrl{-2}       & \qw &        &   & \qw             & \qw \\
    ...\\
      & \lstick{\ket{b_q}} & \qw & \qw             & \qw             & \qw &        &   & \ctrl{-4}       & \qw \\
    }
  \end{tabular}
  \caption{Subroutine circuit $\barr$}
  \label{fig:sub+circuit}
\end{figure}
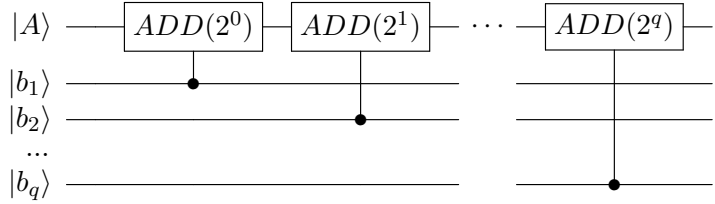

The previous subsection shows that the block $ADD$ has a linear size in the size of the register. While $\ket{A}$ of size $p$ and while there are $q$ blocks, the block is of size $pq$ and needs $p$ ancillary qubits.

\subsection*{Appendix B. Size and complexity of subroutine 1 in variant 2}
\addcontentsline{toc}{subsection}{Appendix B. Size and complexity of subroutine 1 in variant 2}

We detail below the length and the size of the subroutine \textit{Cost calculation} with variant 2 (cf. Table 10). We can assume $n = 2^q$ by adding ($2^q - n$ qubits with $q = \rfloor \log_2 n \lfloor + 1$ if that is not the case.

It is trivial that there are $q = \log_2(2^q) \leq \log_2(2n)$ steps.

We can see that:
\begin{itemize}
  \item $2^q/2^1$ registers are used only for step 1 and store at most 1 cost ($\leq d$). They need to be of size $\lfloor \log_2(d) \rfloor +1$.
  \item $2^q/2^2$ registers are used only for steps 1 and step 2 and store at most 2 cost ($\leq 2d$). They need to be of size $\lfloor \log_2(2d) \rfloor +1 \leq \lfloor \log_2(d) \rfloor + 1 + 1 $.
  \item ...
  \item $2^q/2^q = 1$ register is used only for steps 1 to q and store at most $2^q$ costs ($\leq nd$). It needs to be of size $\lfloor \log_2(2^qd) \rfloor +1 \leq q + \lfloor \log_2(d) \rfloor + 1$.
\end{itemize}
Thus we need a total of:
$$\sum_{k=0}^\infty \lfloor \frac{2^q}{2^{k+1}} \rfloor( \lfloor \log_2(d) \rfloor + k + 1) = 2^{q-1} \sum_{k=0}^\infty \frac{ \lfloor \log_2(d) \rfloor + k + 1}{2^k} \leq 2^{q-1}(\lfloor \log_2(d)\rfloor + 3)
$$

Then we need in overall less than $n(\lfloor \log_2(d) \rfloor + 3) = O(n)$ qubits.

\subsection*{Appendix C. Tables of simulation results}
\addcontentsline{toc}{subsection}{Appendix C. Tables of simulation results}
\subsubsection*{Without penalty}

\begin{table}[h!]
  \centering
  \begin{tabular}{|c|c|c|c|c|c|c|c|}
    \hline
    n /\ P & 2     & 3     & 4     & 5     & 6     & 7     & 8     \\
    \hline
    1      & 0.895 & 0.893 & 0.933 & 0.982 & 0.943 & 0.99  & 0.99  \\
    \hline
    2      & 0.827 & 0.899 & 0.911 & 0.974 & 0.993 & 0.988 & 0.983 \\
    \hline
    3      & 0.857 & 0.896 & 0.958 & 0.959 & 0.967 & 0.989 & 0.983 \\
    \hline
    4      & 0.873 & 0.898 & 0.941 & 0.972 & 0.977 & 0.989 & 0.983 \\
    \hline
    5      & 0.840 & 0.891 & 0.940 & 0.955 & 0.982 & 0.992 & 0.982 \\
    \hline
    6      & 0.857 & 0.922 & 0.945 & 0.950 & 0.978 & 0.985 & 0.987 \\
    \hline
    7      & 0.838 & 0.903 & 0.927 & 0.966 & 0.978 & 0.991 & 0.981 \\
    \hline
    8      & 0.856 & 0.925 & 0.943 & 0.972 & 0.973 & 0.979 & 0.973 \\
    \hline
    9      & 0.849 & 0.920 & 0.934 & 0.964 & 0.979 & 0.995 & 0.987 \\
    \hline
    10     & 0.848 & 0.903 & 0.944 & 0.956 & 0.976 & 0.992 & 0.978 \\
    \hline
    11     & 0.848 & 0.903 & 0.947 & 0.968 & 0.978 & 0.992 & 0.979 \\
    \hline
  \end{tabular}
  \caption{Relaxed Knapsack QAOA without penalty: results}
  \label{fig:resultswithout}
\end{table}{}

\subsubsection*{With penalty}

\begin{table}[h!]
  \centering
  \begin{tabular}{|c|c|c|c|c|c|c|c|c|c|c|}
    \hline
    n / P & 3     & 4     & 5     & 6     & 7     & 8      & 9      & 10     & 11     & 12     \\
    \hline
    2     & 0.893 & 0.917 & 0.953 & 0.968 & 0.965 & 0.969  & 0.976  & 0.961  & 0.980  & 0.97   \\
    \hline
    3     & 0.892 & 0.902 & 0.925 & 0.945 & 0.938 & 0.939  & 0.946  & 0.946  & 0.940  & 0.94   \\
    \hline
    4     & 0.895 & 0.923 & 0.964 & 0.973 & 0.972 & 0.978  & 0.982  & 0.978  & 0.970  & 0.979  \\
    \hline
    5     & 0.892 & 0.920 & 0.956 & 0.961 & 0.963 & 0.961  & 0.969  & 0.963  & 0.957  & 0.971  \\
    \hline
    6     & 0.882 & 0.919 & 0.950 & 0.944 & 0.951 & 0.951  & 0.953  & 0.953  & 0.946  & NO DAT \\
    \hline
    7     & 0.865 & 0.897 & 0.924 & 0.932 & 0.938 & NO DAT & NO DAT & NO DAT & NO DAT & NO DAT \\
    \hline
  \end{tabular}
  \caption{Relaxed Knapsack QAOA with penalty: results.}
  \label{tab:withpenalty}
\end{table}{}

\pagebreak
\subsection*{Acknowledgements}
\addcontentsline{toc}{section}{Acknowledgements}

The authors would like to thank Dr. Ashley Montanaro for his insightful answers during the development of this algorithm.

\addcontentsline{toc}{section}{References}
\printbibliography

\end{document}